\documentclass[useAMS,usenatbib]{mn2e}
\usepackage{graphicx,amssymb,amsmath,times}
\title{Cosmic-Ray proton spectrum below $100$ TeV in the local region}
\author[Thoudam Satyendra]{Thoudam Satyendra\thanks{E-mail: satyend@barc.ernet.in}\\
Astrophysical Sciences Division,\\
Bhabha Atomic Research Centre, Mumbai-400085, India\\}
\begin{document}
\date{}
\pagerange{}
\maketitle
\label{firstpage}
\begin{abstract}
The propagation of cosmic-ray protons in the Galaxy is discussed under the framework of a three dimensional convection-diffusion model. Starting with the assumption of a uniform and continuous distribution of cosmic-ray sources injecting CRs continuously in the Galaxy and by invoking a supernova explosion at various distances from the Earth, it is found that only those sources located within a distance of $\sim 1.5 kpc$ can produce appreciable temporal fluctuations in the CR proton flux observed at the Earth. So, the construction of the local CR proton spectrum is discussed by seperating the contributions of the distant sources from that of the nearby sources. The contribution from the distant sources is treated in the framework of a continuous source distribution model both in space as well as time, but that of the nearby sources in a discrete space-time source model. The study predicts the presence of at least one  old nearby source with a characteristic age of $\sim 10^5 yrs$ located at a distance of $\sim 0.1 kpc$ to explain the observed proton flux below $\sim 100 GeV$.
\end{abstract}
\begin{keywords}
cosmic rays$-$diffusion$-$supernova remnants
\end{keywords}

\section{Introduction}
Considering the similarity of the power supplied by a supernova explosion (SNE) which is approximately $10^{51} erg/30yrs\sim 10^{42}erg/s$ to the power required to maintain the cosmic-ray (CR) energy density in the Galaxy ($\sim 10^{41}erg/s$), it is widely beleived that the majority of CRs upto $\sim 100 TeV$ are accelerated in supernova remnant (SNR) shock waves (Blandford $\&$ Eichler 1987). At least for CR electrons, this hypothesis is strongly supported by the detections of nonthermal X-ray synchrotron emission from SNRs like SN 1006 (Koyama et al.1995), RX J1713.7-3946 (Koyama et al. 1997), Cas A (Allen et al. 1997), IC 443 (Keohane et al. 1997), G374.3-0.5 (Slane et al. 1999) and RX J0852.0-4622 (Slane et al. 2001). But, for CR protons, their acceleration in SNRs is still not fully understood (Berezhko et al.2002).

Several models have been proposed to discuss the CR propagation in the Galaxy. Among these models, for CR electrons the idea of separation of distant and nearby sources can be found in Shen (1970), Shen and Mao (1971), Cowsik and Lee (1979), Atoyan et al. 1995 and Kobayashi et al. 2004. For CR protons and nuclei, this approach had been discussed by Lingenfelter (1969) using a simple model considering only the particle losses due to escape. 

Much later, Erlykin $\&$ Wolfendale (2001a, 2001b) studied the CR spectra at the Earth using Monte Carlo calculations assuming SNRs as the main sources of CRs. In their model, CR nuclei are assumed to be accelerated throughout the SNR lifetime and liberated into the ISM when the acceleration process ceases. They assumed the SN to be distributed stochastically throughout the nearby Galaxy. Their CR propagation model in the Galaxy includes only the diffusive process, neglecting convection due to the Galactic wind, energy and catastrophic losses due to interactions with the ISM.

Strong $\&$ Moskalenko (2001) have studied the influence of the discrete nature of SNRs on the CR protons spectra in the Galaxy using their \begin{scriptsize}GALPROP\end{scriptsize} CR propagation code. Their calculations involved the stochastic nature of SNR events in the Galaxy and a realistic gas distribution for CR interactions.

In this paper, the contributions of the distant and nearby sources to the local CR proton flux are treated seperately. The study considers only the observed local SNRs as the nearby sources unlike in earlier works and the propagation of CR protons in the Galaxy is treated under the framework of a 3-D convection-diffusion model in which the protons undergo convection, diffusion, adiabatic deceleration, ionization and coulomb losses, particle loss due to pion production processes and injection from the sources. The calculations are done mainly for the CR flux in the vicinity of the Sun. For the separate treatment of the distant and nearby sources, CRs from the distant sources are treated in the framework of a continuous source distribution model both in space and time while those from the nearby sources are considered in a discrete (in space and time) source model. Although the model presented here is true for any type of pointlike sources, the calculations are done considering SNRs as the CR sources. 

The study given in this paper starts with the assumption that the CR sources distribution in the Galaxy are spatially uniform and temporally continuous and then, by invoking a supernova explosion at various distances from the Earth, it is found that only those sources located close to the Earth can produce appreciable temporal fluctuations in the local CR proton flux. Knowing this and considering only the observed local SNRs, the study predicts the presence of atleast one old nearby source (which is referred to as the \textit{missing} source in this paper) in order to explain the observed proton flux below $\sim 100 GeV$. 

The plan of this paper is as follows. Section 2 discusses about the propagation of CR protons in the Galaxy as well as the spatial and temporal dependence of their flux from a single source. Section 3 studies their number density variations in the Galaxy and section 4 discusses about the CR proton flux in the Solar vicinity. Finally, section 5 presents the results and discussions about the study presented in this paper.

\section {Propagation of CR protons in the Galaxy}
\subsection{Propagation properties of CR protons}
CRs undergo diffusion in the Galaxy due to scattering either by magnetic field irregularities or by self excited Alfven and hydromagnetic waves. The diffusion coefficient in the Galaxy is generally obtained using the observed secondary to primary B/C ratios. For energies upto around $100 GeV$, it has the form $D(E)\propto E^{\delta}$ with $\delta=0(0.6)$ for $E< 5 GeV(\geq 5 GeV)$ (Kobayashi et al. 2004 and references therein). However, in the $TeV$ region there is a strong argument from the studies of CR anisotropy that a value of $\delta \sim 0.6$ gives an anisotropy value much larger than the observed value of $\sim 10^{-3}$ (Ptuskin 2005, Ambrosio et al. 2003 and references therein). Also, experimental studies on B/C and (secondary species of Fe)/Fe had supported the idea that $\delta$ changes from $0.6-0.3$ in the $GeV-TeV$ region (Furukawa et al. 2003). But, these constraints imposed by the observed anisotropy are not fully well defined because of the presence of solar wind, magnetic field discontinuities in the solar cavity, large scale structure of local magnetic field lines and the measurements of only the CR velocity component perpendicular to the rotation axis of the Earth (Cesarsky 1980). There is also a difficulty related to the proton spectrum in changing $\delta$ from 0.6 to 0.3 since the proton spectrum would then have a break at around $1TeV$ for a power law injection spectrum whereas observations indicate that the proton spectrum follow a continuous single power law throughout the $(10-10^5) GeV$ energy range. Therefore, in this paper the diffusion coefficient is assumed to follow a single power law in the high energy region and takes the form (Atoyan et al. 1995)
\begin{eqnarray}
D(E)= D_0\left(1+\frac{E}{E_0}\right)^{\alpha}
\end{eqnarray}  
where $E_0=3 GeV$, $\alpha=0.6$ and $D_0$ is chosen such that $D(10 GeV)=10^{28} cm^2 s^{-1}$. Eq. (1) is considered to be valid atleast upto $\sim 100 TeV$ which is approximately the maximum energy of accelerated particles by SNR shock waves (Lagage $\&$ Cesarsky 1983). 

While propagating in the Galaxy although high energy CR protons do not suffer from radiative energy losses like the high energy CR electrons, protons with energies less than around $1 GeV$ suffer significant energy losses due to ionization and Coulomb interactions with the interstellar medium (ISM). Their energy loss rate is given by (Mannheim and Schlickeiser 1994)
\begin{eqnarray}
B(E)\approx 1.82\times 10^{-16} n_H \left(\frac{E}{m_p}\right)^{-0.5}\left(1+0.85\frac{n_{HII}}{n_H}\right)\nonumber\\
GeV s^{-1}
\end{eqnarray} 
where $E$ is in GeV, $m_p=0.938 GeV$ is the proton mass energy, $n_H=n_{HI}+2n_{H_2}$ is the ISM hydrogen atom density and $n_{HII}$ is the density of ionised gas in the ISM (Cordes et al. 1991). The atomic hydrogen density $n_{HI}$ is taken from (Gordon $\&$ Burton 1976) and the molecular hydrogen density $n_{H_2}$ from (Bronfman et al. 1988). For the present study, their values are chosen at the position of our Solar system as $n_H=1.11 cm^{-3}$ and $n_{HII}=0.022 cm^{-3}$. In propagation models like the one which is going to be presented here which include a Galactic wind in the direction perpendicular to the Galactic plane, protons with energy greater than $\sim 1 GeV$ lost their energy mainly due to adiabatic cooling if the wind velocity depends on the Galactic height. In addition to this, CRs are also convected away from the Galactic plane due to the Galactic wind. Thus low energy CRs with energy less than $\sim$ 100 GeV are not able to reach the Earth effectively from distant sources because of their slow diffusion since diffusive timescale goes as $t_{diff}\propto 1/D(E)$. Hence, the effects of various energy loss processes and the Galactic wind are important for low energy protons and should be considered while discussing their propagation in the Galaxy.  

\subsection{CRs from a single point-like source}
In the framework of a convection- diffusion model, the propagation of cosmic ray protons in the Galaxy can be represented by the equation  
\begin{eqnarray}
\nabla\cdot(D\nabla N-\textbf{V}N)
+\frac{\partial}{\partial E}\left\lbrace \left( \frac{\nabla\cdot \textbf{V}}{3}E-b\right) N \right\rbrace -\frac{N}{\tau_{pp}}\nonumber\\
+Q=\frac{\partial N}{\partial t}
\end{eqnarray}
where $E$ is the proton kinetic energy in GeV, $N(\textbf{r},E)$ is the differential number density , $D(\textbf{r},E)=D(E)$ is the diffusion coefficient which is assumed to be spatially uniform, $Q(\textbf{r},E,t)$ is the proton production rate, i.e.  $Q(\textbf{r},E,t)d^3rdEdt$ is the number of protons produced by the source in a volume element $d^3r=dxdydz$ in the energy range $(E,E+dE)$ in time $dt$, $\textbf{V}(\textbf{r})=3V_0z\hat{z}$ is the convection velocity in the direction perpendicular to the Galactic plane with $V_0=15$ km s$^{-1}$ kpc$^{-1}$ and $b(\textbf{r},E) = -B(E)$ is the proton energy loss rate due to ionization and Coulomb interactions which is assumed to be spatially constant. $[-N_p(E,\textbf{r})/\tau_{pp}(E,\textbf{r})]$, $\tau_{pp}=E/(dE/dt)_\pi$ and ${(dE/dt)}_\pi=8\times 10^{-16} n_H E$ in GeV s$^{-1}$ are the catastrophic loss term, energy loss time scale and  energy loss rate respectively due to pion production processes. With these substitutions, Eq. (3) can be simplified in rectangular coordinates as
\begin{eqnarray}
\frac{\partial N}{\partial t}-D(E)\left\lbrace \frac{\partial^2 N}{\partial^2 x}+\frac{\partial^2 N}{\partial^2 y}+\frac{\partial^2 N}{\partial^2 z}\right\rbrace+3V_0\frac{\partial}{\partial z}(zN)-\nonumber\\
\frac{\partial}{\partial E}\left\lbrace \left( V_0E-b\right)N \right\rbrace+\frac{N}{\tau_{pp}}=Q(x,y,z,E,t)
\end{eqnarray} 
The Green's function $G(x,x^{'},y,y^{'},z,z^{'},E,E^{'},t,t^{'})$ of Eq. (4), i.e. the solution for a $\delta$-function type source term $Q(x,y,z,E,t)=\delta(x-x^{'})\delta(y-y^{'})\delta(z-z^{'})\delta(E-E^{'})\delta(t-t^{'})$ is obtained as 
\begin{eqnarray}
G(x,x^{'},y,y^{'},z,z^{'},E,E^{'},t,t^{'})=\frac{A(E)}{8\pi^{3/2}P(E)}\qquad\qquad\quad\nonumber\\
\times {\left[\int^{E{'}}_E \frac{D(u)du}{P(u)}\right]}^{-1}
{\left[\int^{E{'}}_E \frac{D(u)A^2(u)du}{P(u)}\right]}^{-1/2}\nonumber\\
\times exp\left[A_{pp}\int^E_{E^{'}} \frac{du}{P(u)}\right]
exp\left[\frac{-((x^{'}-x)^2+(y^{'}-y)^2)}{4\int^{E^{'}}_E\frac{D(u)du}{P(u)}}\right]\nonumber\\
\times exp\left[\frac{-\left(3V_0z^{'}\int^E_{E^{'}}\frac{A(u)du}{A(E)P(u)}-z^{'}+z\right)^2}{4\int^{E^{'}}_E
\frac{D(u)A^2(u)du}{A^2(E)P(u)}}\right]\nonumber\\
\times \delta\left[t-t^{'}+\int^E_{E^{'}}\frac{du}{P(u)}\right]
\end{eqnarray} 
where $P(E)=V_0E+B(E)$, $A_{pp}=8\times 10^{-16} n_H$ and
\begin{equation}
A(E)=exp\left[3V_0\int^E\frac{du}{V_0u+B(u)}\right]\nonumber
\end{equation} 
Then, the general solution of Eq. (4) is given by
\begin{eqnarray}
N(x,y,z,E,t)=\int^{\infty}_{-\infty}dx^{'}\int^{\infty}_{-\infty}dy^{'}\int^{\infty}_{-\infty}dz^{'}\int^{\infty}_EdE^{'}\nonumber\\
\int^t_{-\infty}dt^{'}Q(x^{'},y^{'},z^{'},E^{'},t^{'})\nonumber\\
\times G(x,x^{'},y,y^{'},z,z^{'},E,E^{'},t,t{'})
\end{eqnarray} 
Now, to calculate the number density $N(x,y,z,E,t)$ due to a point source located at $(0,0,0)$ producing $q(E)$ protons of energy $E$ per unit energy interval at the time of explosion, let us start by considering a small source volume $\rho$ bounded by $|x|=x_1, |y|=y_1, |z|=z_1$ where protons are generated. Before the explosion protons are assumed to be distributed uniformly in the source volume with density $q_0(E)$. In other words we can write 
\begin{eqnarray}
Q(x,y,z,E,t)=q_0(E)q(t),\qquad inside\;the\;source,\nonumber\\
=0,\qquad\qquad\quad\; outside\;the\;source, 
\end{eqnarray} 
and for a point source\\ 
\\
lim   $q_0(E)\;\rho=q(E)$\\
$\rho \rightarrow 0$\\
\\
With this assumption the proton density outside the point source positioned at $(0,0,0)$ can be obtained by introducing Eqs. (5)$\&$(7) in (6) as
\begin{eqnarray}
N(x,y,z,E,t)=\frac{A(E)}{8\pi^{3/2}P(E)}\int^t_{-\infty}dt^{'}q(t^{'})q(U)P(U)\nonumber\\
\times {\left[\int^U_E \frac{D(u)du}{P(u)}\right]}^{-1}
{\left[\int^U_E \frac{D(u)A^2(u)du}{P(u)}\right]}^{-1/2}\nonumber\\
\times exp\left[A_{pp}(t^{'}-t)\right]
exp\left[\frac{-(x^2+y^2)}{4\int^U_E\frac{D(u)du}{P(u)}}\right]\nonumber\\
\times exp\left[\frac{-z^2}{4\int^U_E
\frac{D(u)A^2(u)du}{A^2(E)P(u)}}\right]
\end{eqnarray}  
where 
\begin{eqnarray}
U=\left[\frac{exp\left\lbrace\frac{3V_0(t-t^{'})}{2}\right\rbrace\left\lbrace V_0E^{3/2}+C\right\rbrace-C}{V_0}\right]^{2/3}\nonumber
\end{eqnarray}  
and \begin{eqnarray}
C=1.82\times 10^{-16} n_H \sqrt{m_p}\left(1+0.85\frac{n_{HII}}{n_H}\right)\nonumber
\end{eqnarray} 
Assuming burstlike injection of CR protons from the source exploding at time $t=t_0$, we can write $q(t^{'})=\delta (t^{'}-t_0)$ which gives
\begin{eqnarray}
N(x,y,z,E,t)=\frac{A(E)}{8\pi^{3/2}P(E)}q(U_0)P(U_0)exp\left[A_{pp}(t_0-t)\right]\nonumber\\
\times {\left[\int^{U_0}_E \frac{D(u)du}{P(u)}\right]}^{-1}
{\left[\int^{U_0}_E \frac{D(u)A^2(u)du}{P(u)}\right]}^{-1/2}\nonumber\\
\times exp\left[\frac{-(x^2+y^2)}{4\int^{U_0}_E\frac{D(u)du}{P(u)}}\right]exp\left[\frac{-z^2}{4\int^{U_0}_E\frac{D(u)A^2(u)du}{A^2(E)P(u)}}\right]
\end{eqnarray} 
with
\begin{eqnarray}
U_0=\left[\frac{exp\left\lbrace\frac{3V_0(t-t_0)}{2}\right\rbrace\left\lbrace V_0E^{3/2}+C\right\rbrace-C}{V_0}\right]^{2/3}\nonumber
\end{eqnarray}
In cylindrical co-ordinates $(r,\theta ,z)$, Eq. (9) becomes
\begin{eqnarray}
N(r,\theta ,z,E,t)=\frac{A(E)}{8\pi^{3/2}P(E)}q(U_0)P(U_0)exp\left[A_{pp}(t_0-t)\right]\nonumber\\
\times {\left[\int^{U_0}_E \frac{D(u)du}{P(u)}\right]}^{-1}
{\left[\int^{U_0}_E \frac{D(u)A^2(u)du}{P(u)}\right]}^{-1/2}\nonumber\\
\times exp\left[\frac{-r^2}{4\int^{U_0}_E\frac{D(u)du}{P(u)}}\right]exp\left[\frac{-z^2}{4\int^{U_0}_E\frac{D(u)A^2(u)du}{A^2(E)P(u)}}\right]
\end{eqnarray} 
It should be mentioned here that the proton source spectrum $q(E)$ is taken as
\begin{equation}
q(E)=k(E^2+2Em_p)^{-(\Gamma+1)/2}\;(E+m_p)
\end{equation} 
The constant $k$ is obtained by normalizing the proton source spectrum such that the total amount of kinetic energy contained in the injected protons is $0.2$ times the total supernova explosion energy.
\begin{figure}
\centering
\includegraphics*[width=0.31\textwidth,angle=270,clip]{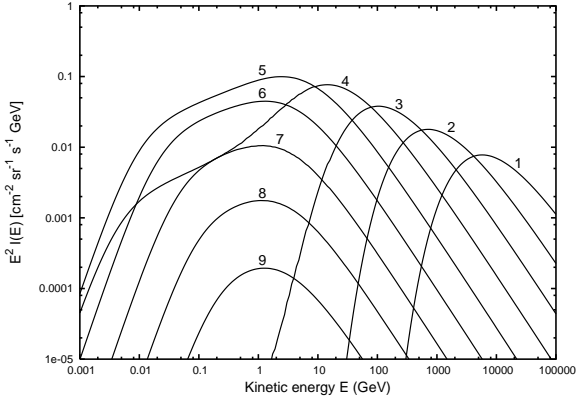}
\caption{\label {fig1} Temporal behaviour of CR protons at a distance $r=0.1 kpc$, $z=0$ from a pointlike source. The spectra numbered as $(1,2,3,4,5,6,7,8,9)$ are calculated at times $t=(10^3,3\times 10^3,10^4,3\times 10^4, 10^5,3\times 10^5, 10^6, 3\times 10^6,10^7) yrs$ respectively. The calculations are done for total output source energy $W=10^{51}ergs, \Gamma=2.4, \alpha=0.6, D(10 GeV)=10^{28} cm^2s^{-1}$ and $V_0=15km s^{-1} kpc^{-1}$ assuming a burstlike injection of the protons at time $t=0$. }
\end{figure}

Fig. 1 shows the $E^2$ times the proton flux $I(E)=(c/4\pi)N(E)$ at a distance $r=0.1 kpc$, $z=0$ from a source at different times $t$ after their burstlike injection into the ISM at time $t=0$. It can be seen that the energy spectra are modified considerably at different times due to the combined effects of different propagation and energy loss rates of protons at different energies. At shorter times $t\lesssim 10^4 yrs$, only the high energy particles are able to travel the distance because of their faster diffusion in the Galaxy. Low energy particles are not able to cover the distance in the short time. But at longer times $t\gtrsim 10^4 yrs$, the low energy particles also start reaching the observer and by that time the high energy particles has already passed by and hence, their flux is significant suppressed. At still longer times $t\gtrsim 10^6 yrs$, even the low energy protons have already passed by and the particle fluxes are found to decrease with time at all the energies. 

The particle fluxes are also found to depend strongly on the source distance $r$. The distance dependence is found to follow a Gaussian function (see Eq. 10). Since the CR sources are distributed within a region very closed to the Galactic plane (half thickness $\sim 150 pc$) and also our Solar system is only $15 pc$ away from the plane (Cohen 1995), it will be assumed from now on that the sources as well as the Sun are positioned at $z=0$. Fig. 2 shows the various CR proton flux in the solar vicinity due to a source located at various distances from the Sun. The maximum flux of protons of energy $E$ from a source after their injection into the ISM at $t=0$ would be observed at time $t_m(E)=r^2/6D(E)$. The fluxes shown in the figure are calculated for $r=(0.05,0.1,0.5,1.0,1.5)kpc$ at their respective $t_m$'s for proton energy $E=10 GeV$. The magnitude of the maximum flux at $10GeV$ is found to decrease with the source distance $r$. The calculation includes the effects of Solar modulation using the force field approximation (Gleeson $\&$ Axford 1968) with modulation parameter $\Phi =400 MV$.
\begin{figure}
\centering
\includegraphics*[width=0.31\textwidth,angle=270,clip]{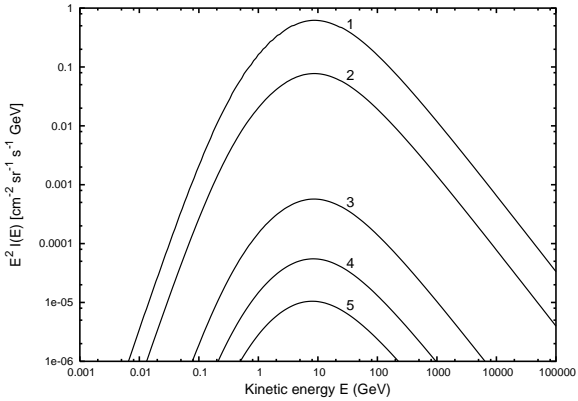}
\caption{\label {fig2} Various CR proton flux at the position of the Sun due to a single source located at various distances $r$. The spectra marked as $(1,2,3,4,5)$ are calculated for source distances $r=(0.05,0.1,0.5,1.0,1.5) kpc$ at their respective times $t_m(E)=r^2/6D(E)$ for proton energy $E=10 GeV$, where $t_m$ is defined as the time at which the maximum flux of CR protons of energy $E$ is observed after their injection from the sources at time $t=0$. The model parameters are the same as in Fig.1. The Solar modulation parameter $\Phi =400 MV$.}
\end{figure}

\subsection{CRs from the source distribution in the Galactic disk}
The propagation of CR protons in the Galaxy in the case of stationary injection from a continuous and uniform source distribution in the Galactic disk can be represented by the equation 
\begin{eqnarray}
\nabla\cdot(D\nabla N-\textbf{V}N)
+\frac{\partial}{\partial E}\left\lbrace \left( \frac{\nabla\cdot \textbf{V}}{3}E-b\right) N \right\rbrace -\frac{N}{\tau_{pp}}\nonumber\\
+Q=0
\end{eqnarray}
where the symbols have the same representation as in section 2.2. The proton density at a given location in the Galactic disk due to all the sources located beyond a distance $r$ from the point is obtained by solving Eq. (12) assuming a uniform and continuous distribution of sources in the disk and is given by (Thoudam 2006)
\begin{eqnarray}
N(E)=\frac{k^{'}A(E)}{2\sqrt{\pi}P(E)}\int^{\infty}_E dE^{'}q(E^{'})exp\left[A_{pp}\int^E_{E^{'}} \frac{du}{P(u)}\right]\nonumber\\
\times{\left[\int^{E{'}}_E \frac{D(u)A^2(u)du}{P(u)}\right]}^{-1/2}exp\left[\frac{-r^2}{4\int^{E^{'}}_E\frac{D(u)du}{P(u)}}\right]
\end{eqnarray} 
Here, the source spectrum $q(E)$ is taken as the same form as in Eq. (11) with normalization $k^{'}$ (in place of $k$) which is obtained assuming a Galactic (SNE) rate of 25 SNE $Myr^{-1}kpc^{-2}$ in the disk (Grenier 2000) and 0.2 times the total SNE energy going into the proton kinetic energy. It should be noted that in the case of a stationary injection of CRs into the Galaxy from sources continuously distributed in the Galactic disk, the equilibrium CR proton spectrum in the disk depends weakly on the actual geometry of the source distribution for a constant CR energy density in the Galaxy.

\section {CR proton density variations in the Galaxy}
Let us try to examine the sources which can produce appreciable temporal density fluctuations of CR protons at a given location in the Galaxy. The study starts with the assumption of a uniform and continuous distribution of cosmic-ray sources injecting CRs continuously in the Galaxy. These sources will contribute the background CRs in the study. Then, let us try to invoke a SNE at increasing distances from the point and see its effects on the CR density at the point. 

Figs. 3$\&$4 shows the temporal variation of the proton density at $10 GeV$ and $10 TeV$ energies respectively for a SNE at $r=(0.1,0.5,1,1.5)kpc$ distances from the point. The calculation assumes a particle release time of $t=0$ from the source. It can be seen that the temporal density variation decreases as the distance of the SNR increases. For $10GeV$ protons, the amplitude of variation decreases from $\sim 90\%$ at $r=0.1 kpc$ to $\sim 0.01\%$ at $r=1.5kpc$. Similarly for $10TeV$ protons, the amplitude varies as $\sim 700\%$ to $\sim 0.2\%$ as $d$ increases from $0.1kpc$ to $1.5kpc$. The higher amplitude in the case of $10 TeV$ protons is expected because of the energy dependence of the diffusion coefficient which allow higher energy particles to diffuse faster than the low energy ones in the Galaxy. With this study, it can be concluded that the propagation of CRs from nearby sources (say within $\sim 1.5kpc$) should be treated in a time dependent model whereas those from far sources (those beyond a distance of $\sim 1.5 kpc$) can be discussed in a steady state model. Also, the spatial treatment between the nearby and distant sources should be different since the mean distance between the sources can be considered to be comparable to their distances from the Earth for nearby sources whereas that of the distant sources can be assumed to be very much less than their distances from us. So, the models going to be adopted are the following :

$(1)$ For nearby sources $(r$$<$$1.5kpc)$ : discrete both in space and time  

$(2)$ For distant sources $(r$$\geq$$1.5kpc)$: continuous both in space and time

\section {CR protons in the solar vicinity}
Fig. 5 shows the model CR proton flux in the solar vicinity together with the observed data. The model spectra comprises of two components. One component (the distant component) is contributed by distant sources located beyond a distance of $\sim 1.5 kpc$ from us and the other component (the local component) is from nearby known SNRs located within a distance of $1.5 kpc$. The two  components are treated seperately as discussed in the last section. 
\begin{figure}
\centering
\includegraphics*[width=0.31\textwidth,angle=270,clip]{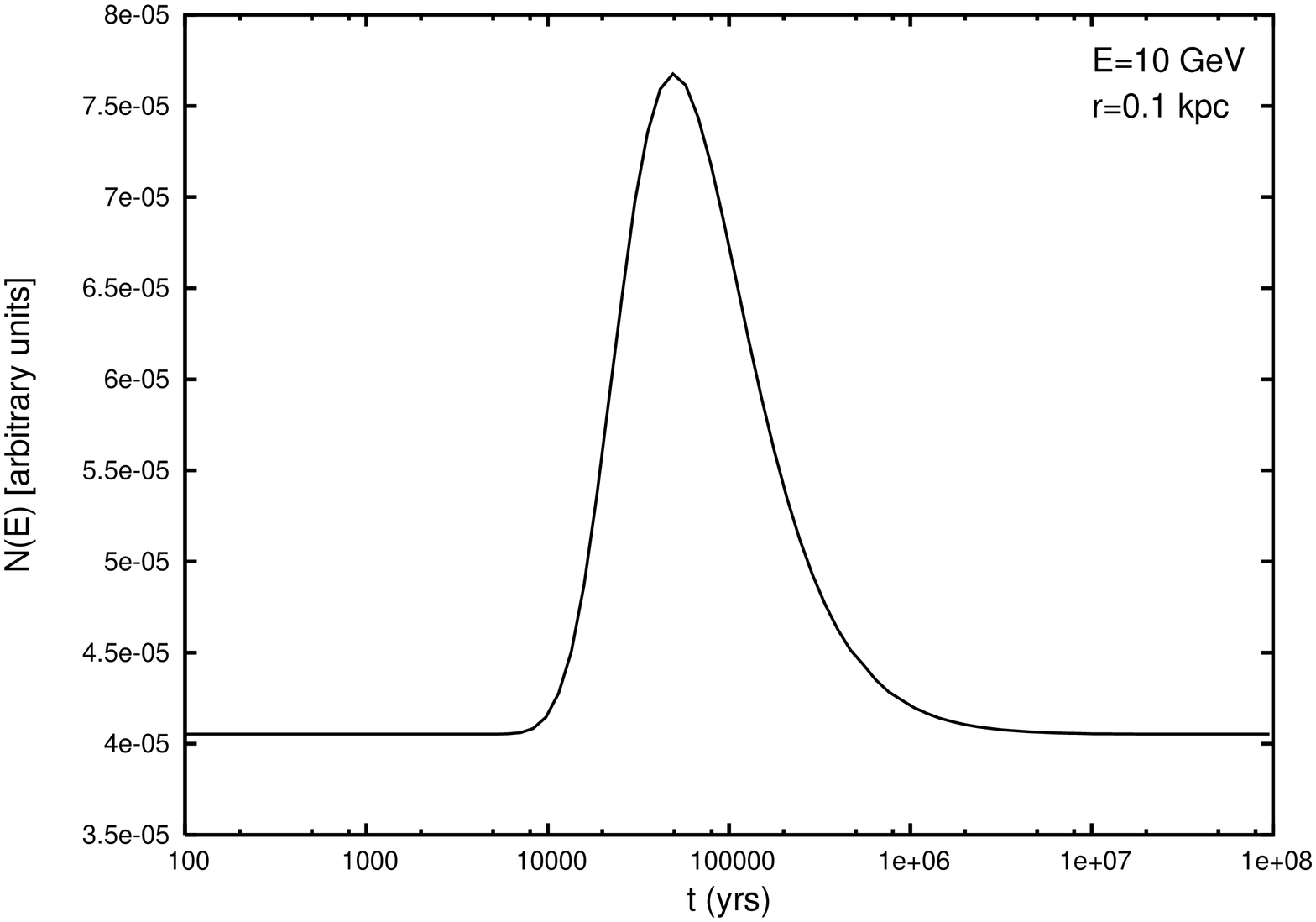}
\includegraphics*[width=0.31\textwidth,angle=270,clip]{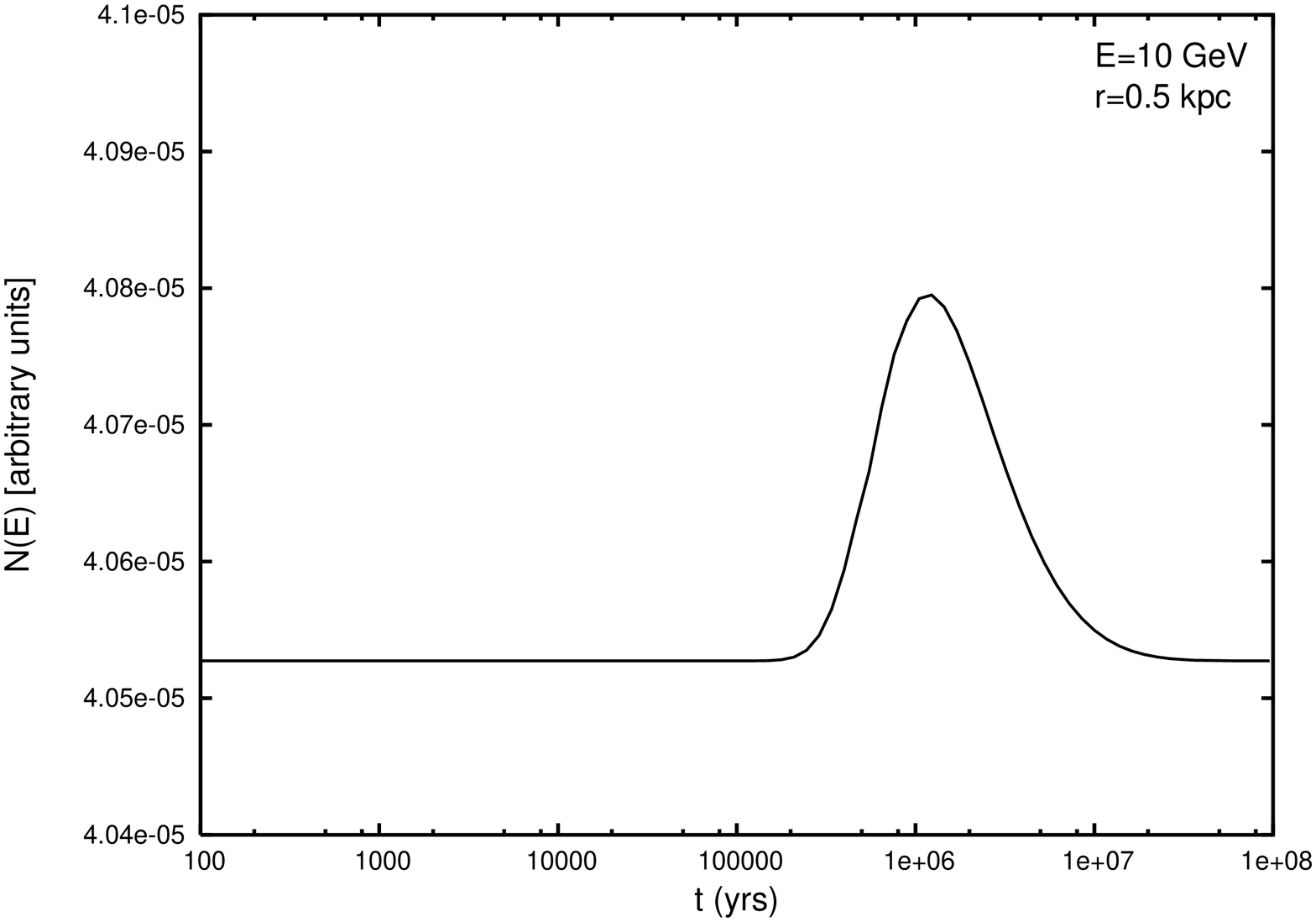}
\includegraphics*[width=0.31\textwidth,angle=270,clip]{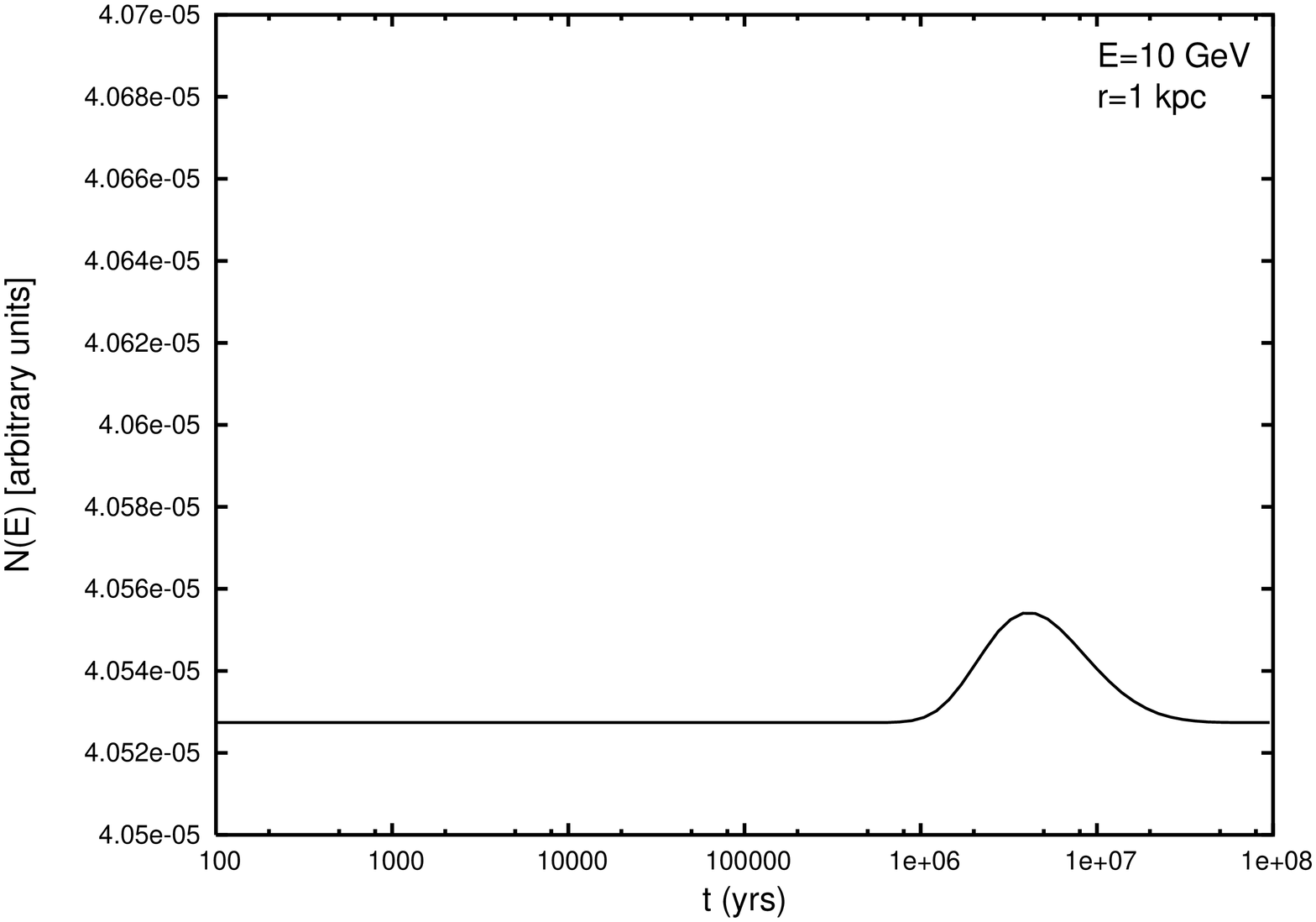}
\includegraphics*[width=0.31\textwidth,angle=270,clip]{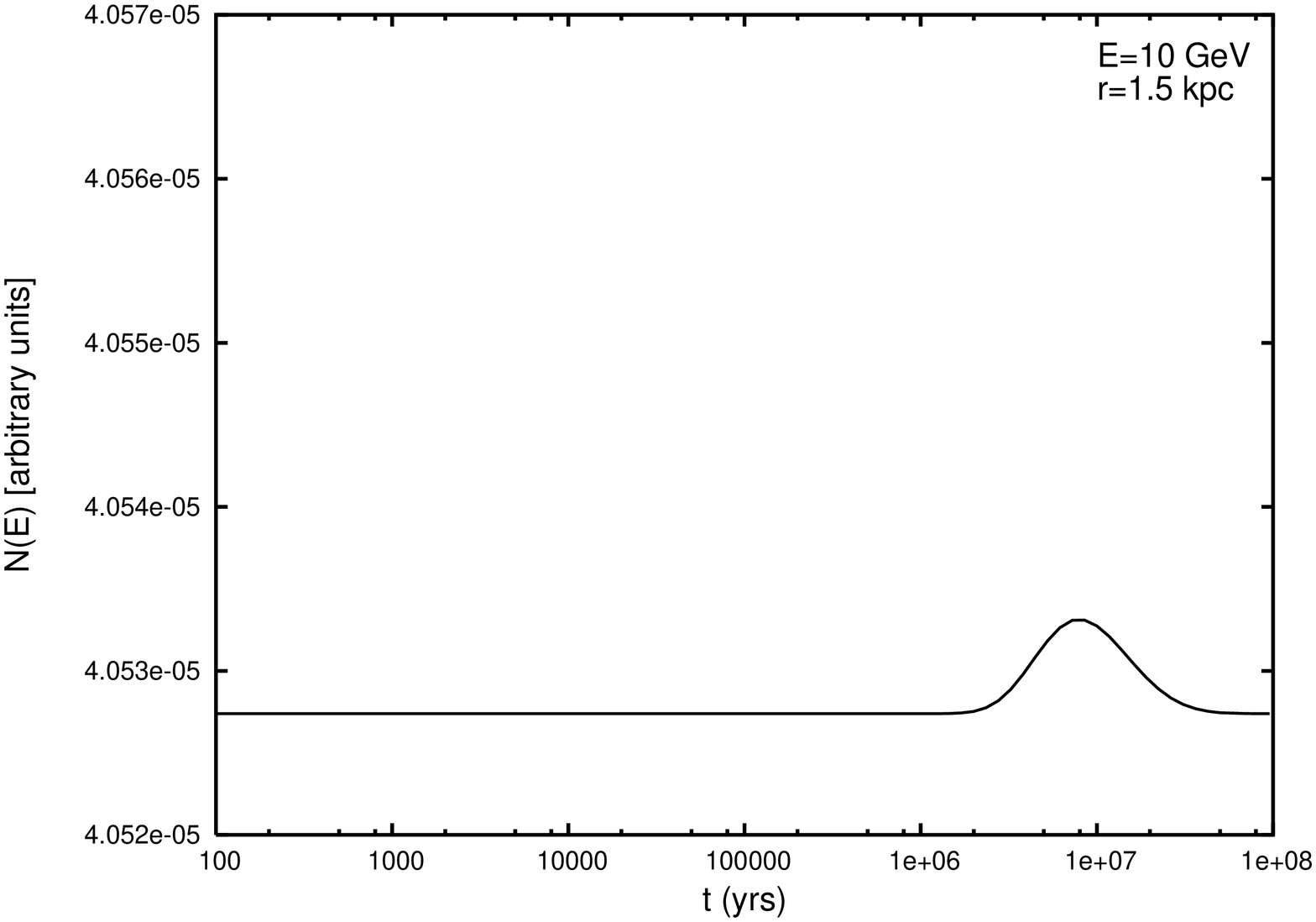}
\caption{\label {fig2} Temporal density variations of $10 GeV$ CR proton at a given location in the Galactic disk due to a source located at distances $r=(0.1,0.5,1,1.5)kpc$ (\textit{top} to \textit{bottom}) from the point. The calculation is for a particle release time of $t=0$ from the source.}
\end{figure}
\begin{figure}
\centering
\includegraphics*[width=0.31\textwidth,angle=270,clip]{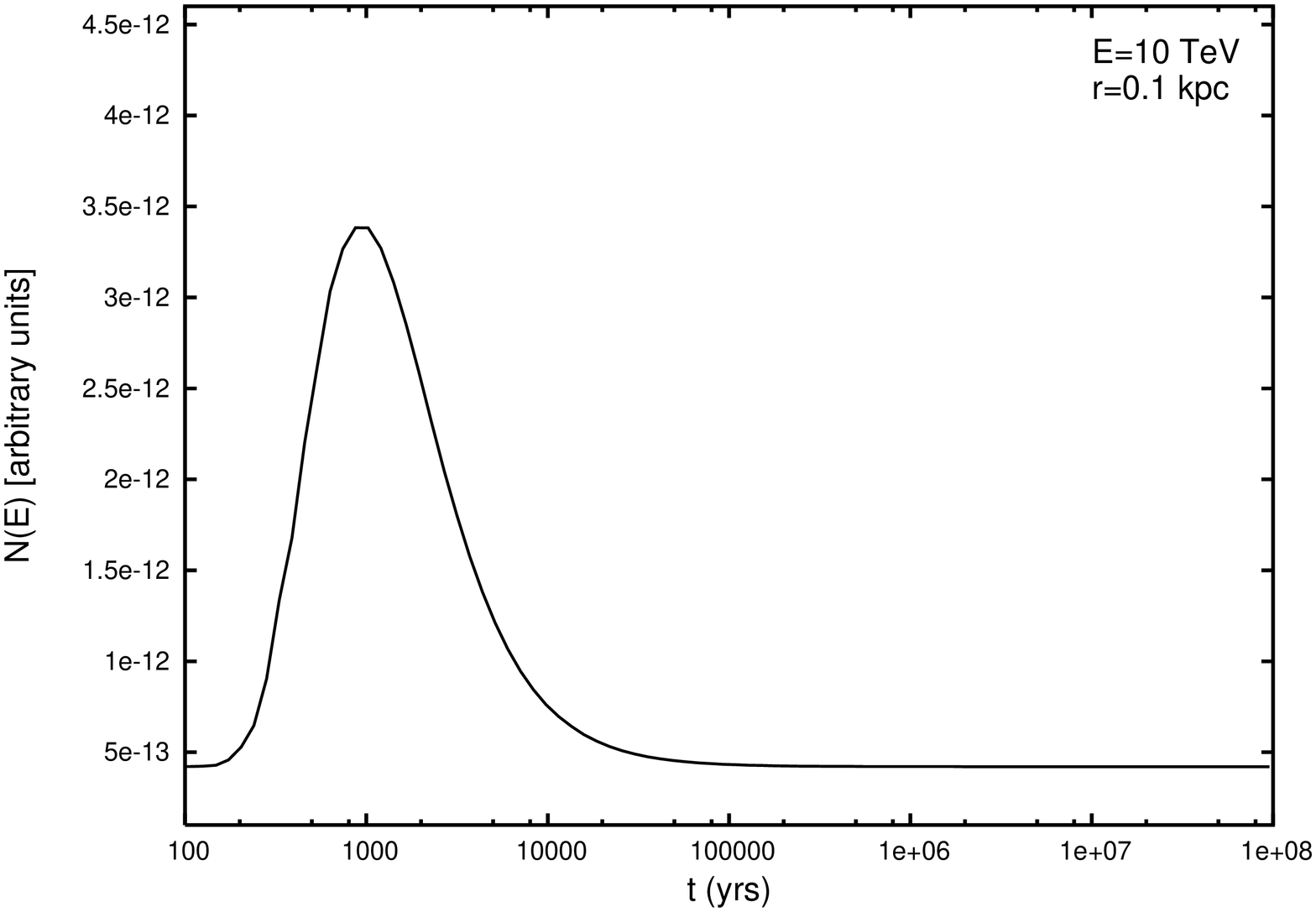}
\includegraphics*[width=0.31\textwidth,angle=270,clip]{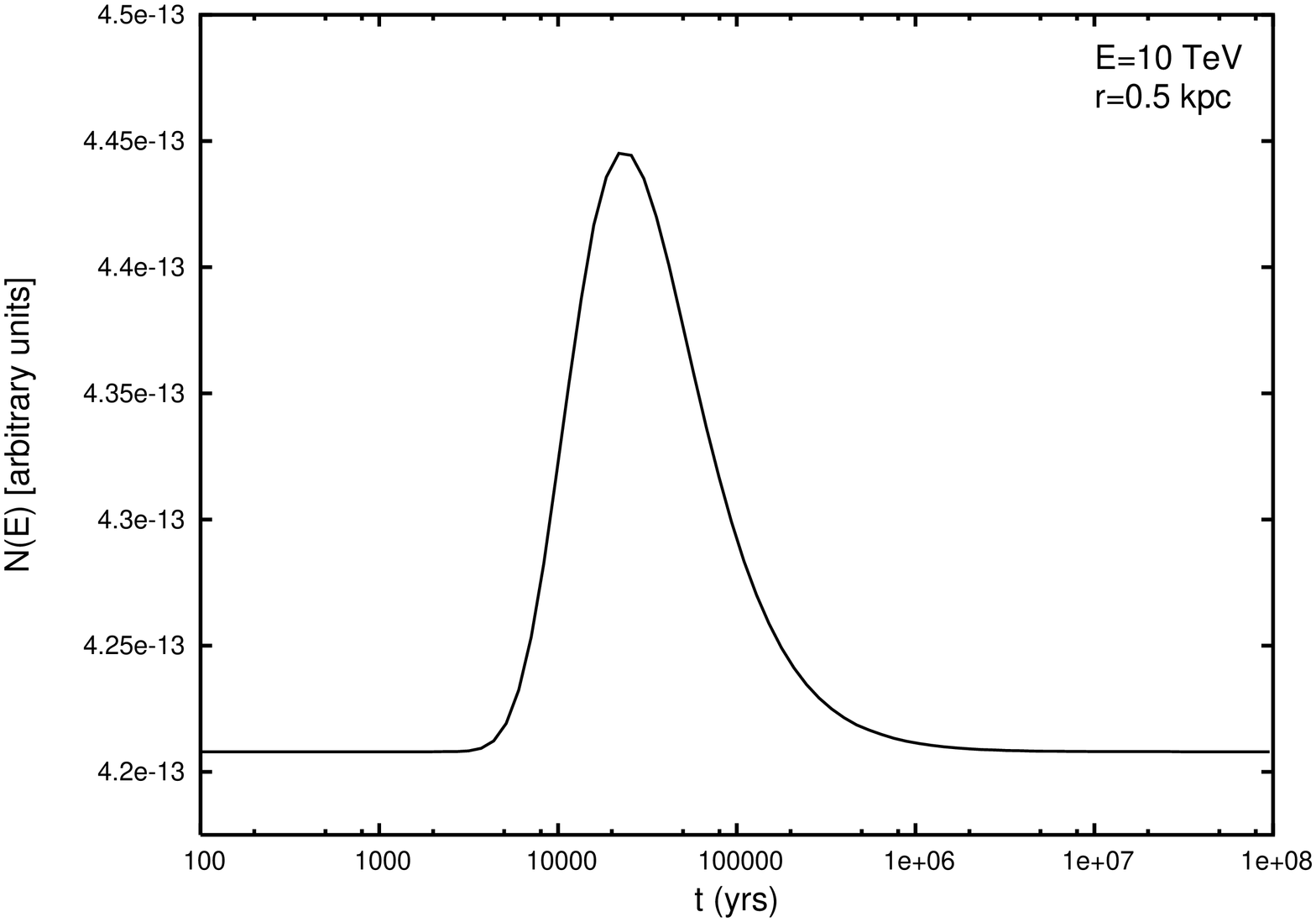}
\includegraphics*[width=0.31\textwidth,angle=270,clip]{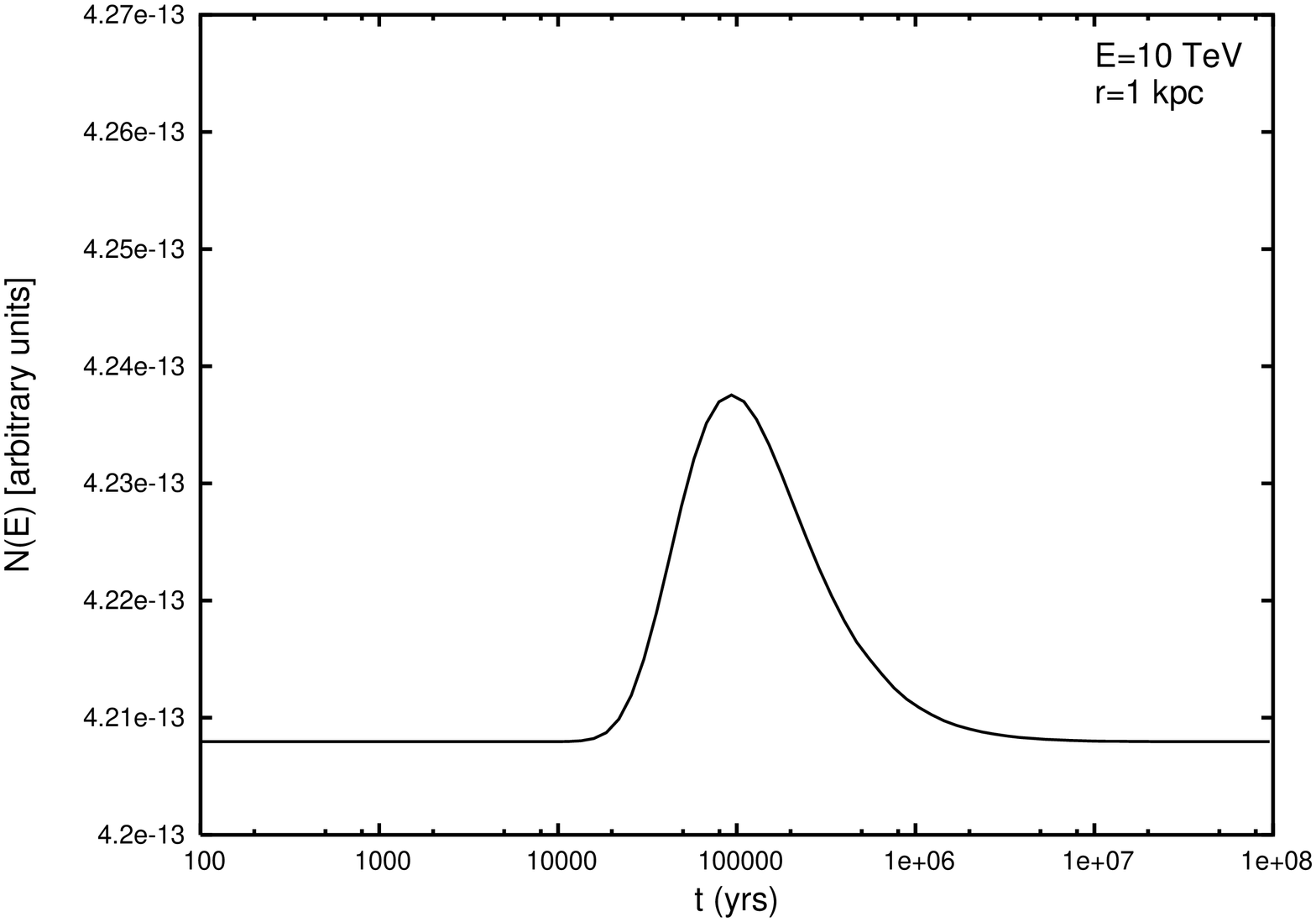}
\includegraphics*[width=0.31\textwidth,angle=270,clip]{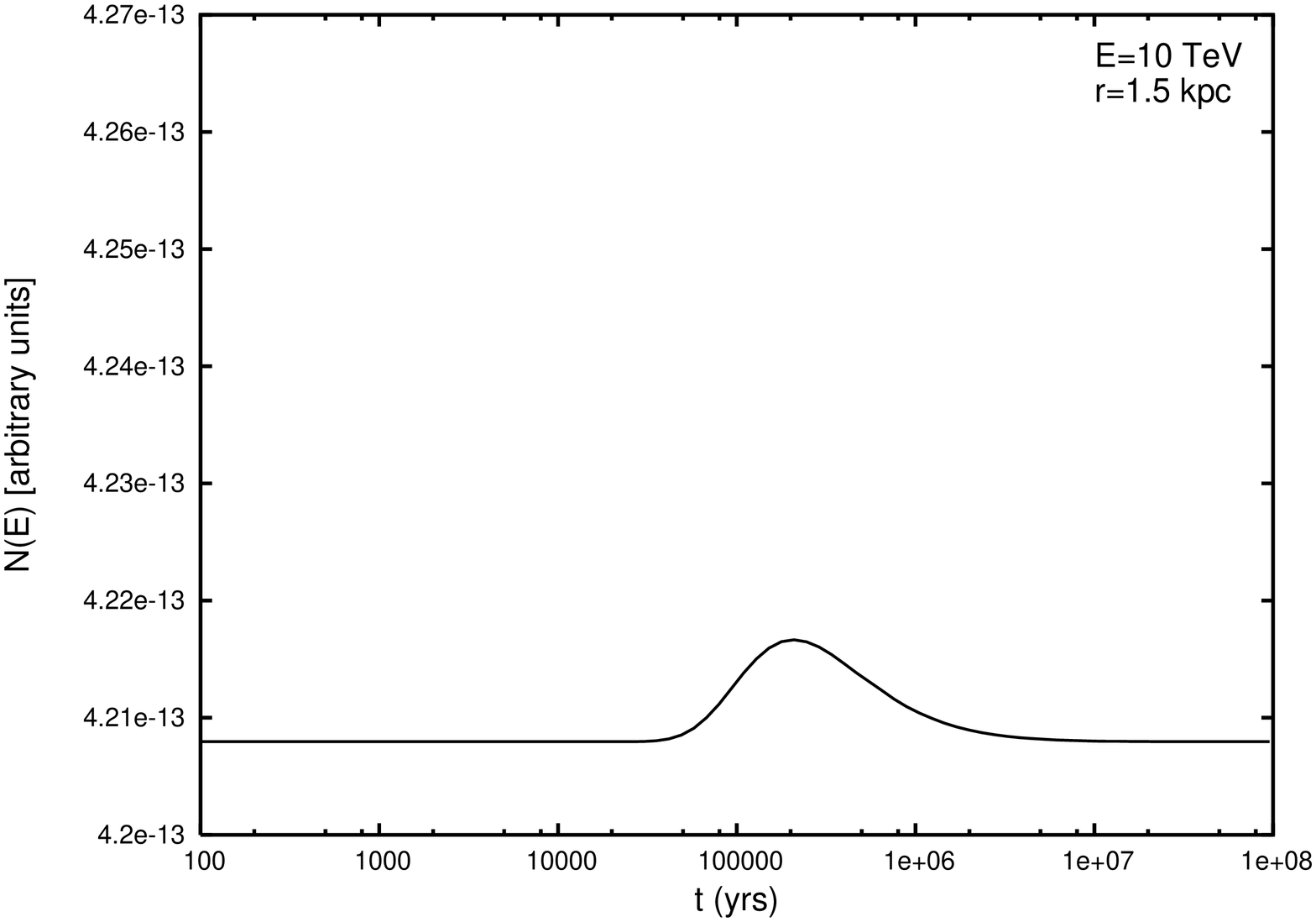}
\caption{\label {fig3} Same as in Fig. 3 but for a proton energy of $10 TeV$.}
\end{figure}

From a compilation of several parameters of 231 all-type SNRs in the Galaxy based on radio observations as given in Table 1 of Jian-Wen et al. (2005), it is found that there are only 11 SNRs which are located within a distance of $1.5 kpc$ from the Sun. Out of these, 9 SNRs are with  known ages. The two SNRs whose ages are not known are G311.5-0.3 and Lupus Loop, each located at a distance of $\sim 1.2 kpc$ from us. In addition to these SNRs detected in radio, there are also other SNRs within the region which are detected in other wavelength bands like Geminga and Monogem. The Geminga supernova is found to emit only X-rays and gamma-rays but not radio waves. Monogem is visible only in X-rays. The distances from the Sun $(r)$ and the ages $(t)$ of the nearby SNRs with distances $r<1.5 kpc$ whose ages are known are listed down in Table 1. The SNR RXJ1713.7-3946 is not included in the list because its distance is still not exactly known. The distance to the SNR is $(6\pm 1) kpc$ in contrast to the distance of $1 kpc$ estimated from soft X-ray obsorption (Koyama et al. 1997). Its age is approximately $10^4 yrs$ for a distance of $r=6 kpc$ or $2\times 10^3 yrs$ for $r=1 kpc$. 
\begin{table}
\centering
\caption{Parameters of SNRs with known ages located within a distance of $1.5 kpc$ from the Sun :}
\begin{tabular}{@{}llrrlrlr@{}}
\hline
SNR name & distance $r$ && Age $t$ \\
         & $(kpc)$      && (yrs)\\
\hline
G65.3+5.7   &     1.0  &&   14000\\
G73.9+0.9   &     1.3  &&   10000\\
Cygnus Loop &     0.4  &&   14000\\
HB21        &     0.8  &&   19000\\
G114.3+0.3  &     0.7  &&   41000\\
CTA1        &     1.4  &&   24500\\
HB9         &     1.0  &&   7700 \\
Vela        &     0.3  &&   11000 \\
G299.2-2.9  &     0.5  &&   5000\\
Monogem\footnote     &     0.3  &&   86000\\
Geminga\footnote     &     0.4  &&   340000\\
\hline
$^1$ Plucinsky et al. 1996;\\
$^2$ caraveo et al. 1996; 
\end{tabular}
\end{table}
\begin{figure}
\centering
\includegraphics*[width=0.31\textwidth,angle=270,clip]{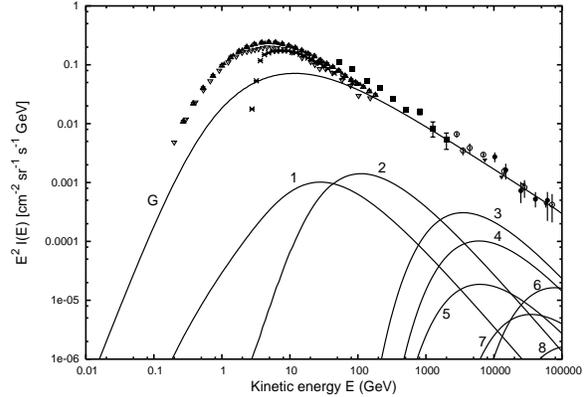}
\caption{\label {fig5} The flux of CR protons near the Sun produced by distant sources and nearby known SNRs listed in Table 1. The spectrum marked as $G$ is the contribution from distant sources located beyond $1.5 kpc$ from us and the spectra numbered as $(1,2,3,4,5,6,7,8)$ are the contributions from individual SNRs with $1-$ Geminga, $2-$ Monogem, $3-$ Vela, $4-$ Cygnus Loop, $5-$ G114.3+0.3, $6-$ G299.2-2.9, $7-$ HB21 and $8-$ G65.3+5.7. The calculations are done for total output source energy $W=10^{51}ergs, \Gamma=2.45, \alpha=0.6, D(10 GeV)=10^{28} cm^2s^{-1}$ and $V_0=15km s^{-1} kpc^{-1}$ assuming a burstlike injection of the protons at time $t=0$. The Solar modulation parameter $\Phi =400 MV$. Data points shown corresepond to the local fluxes of CR protons measured by different groups (see Thoudam 2006 for details).}
\end{figure}

The calculation of the distant component is done for a SNE rate of two times the Galactic value. It can be seen from the figure that non of the observed nearby SNRs nor the distant sources are  able to contribute the observed data effectively below $\sim 100 GeV$. It is because for low energy protons, their diffusion time from distant sources becomes comparable to their ionization, coulomb and adiabatic energy loss timescales due to which their fluxes are significantly suppressed. Also they may be convected away from the Galactic plane before reaching us due to the presence of Galactic wind. Only high energy protons which diffuse faster can reach us from distant sources before they are lost in the intervening ISM. On the other hand, since the fluxes from nearby SNRs depend strongly on their distances $r$ as well as their ages $t$, the low energy protons from the knwon nearby sources have not yet reached us effectively except those from Monogem and Geminga. But, below $100 GeV$ the contributions of these two SNRs are approximately one to two orders of magnitude less than the observed data. Above $100 GeV$ upto the maximum energy considered here $(100 TeV)$, most of the contributions are from the distant sources. Among the nearby SNRs, only the Vela and Cygnus Loop are found to contribute the maximum flux in the high energy region. However, their contributions are around an order of magnitude less than the observed flux. 

To explain the observed proton spectrum below $100GeV$, the presence of atleast one old SNR with an approximate age of $t\sim 10^5 yrs$ located at a distance of $r\sim 0.1kpc$ is essential. The basic parameters $(r,t)$ of the \textit{missing} SNR is possible to be determined if the particle spectra at different times (as shown in Fig.1) are carefully examined together with the variation of the maximum flux with $r$ (shown in Fig.2). Since the maximum flux depends strongly on the distance of the source $r$ from the Sun, the \textit{missing} source cannot be located far beyond $\sim 0.1kpc$ or nearer than it otherwise the proton flux will be either below or above the observed flux. Also, since the proton spectrum depends strongly on time $t$, the missing SNR cannot be younger nor older than $\sim 10^5 yrs$. In short, there is only one set of possible combination of $(r,t)$ for the \textit{missing} source which can explain the observed the proton spectrum below $100 GeV$.
\begin{figure}
\centering
\includegraphics*[width=0.31\textwidth,angle=270,clip]{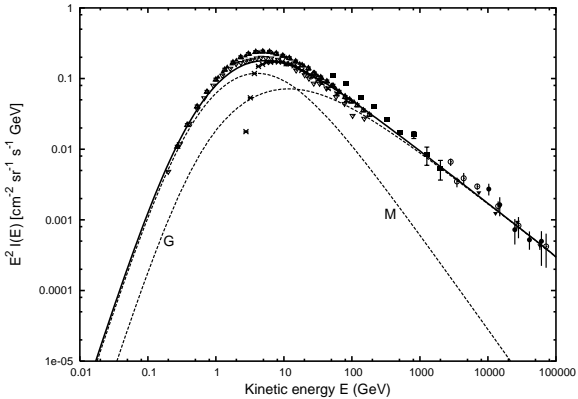}
\caption{\label {fig6} Total CR proton spectrum in the Solar vicinity. The thick solid line represents the total model spectrum which consists of the contribution from distant sources located beyond $1.5 kpc$ from us (marked as G) and the contribution from a \textit{missing} source (probably SNR) having a characteristic age of $t=10^5 yrs$ located at a distance of $r=0.09 kpc$ (marked as M). The calculation is done assuming burst-like injection of CR protons from the source at time $t=0$. All the model parameters and the proton data points are the same as in Fig. 5}
\end{figure}

Fig. 6 shows the predicted total local CR proton spectrum which consists of the contributions from distant sources located beyond $1.5 kpc$ and that from an SNR with a characteristic age of $t=10^5 yrs$ positioned at a distance of $r=0.09kpc$ from the Sun. The total spectrum does not include the contributions from the observed SNRs listed in Table 1 since their fluxes are found to be almost one-two orders of magnitude less than the observed flux. It can be seen that now the observed flux is explained effectively with the inclusion of the \textit{missing} source.

\section {Results and discussions}
The model presented in the paper involves the seperate treatment of the distant and nearby sources for CR protons near the Sun. The distant sources are treated in a model continuous both in space and time whereas the nearby sources are discussed in a discrete model in space as well as time. The study shows that the total contribution of the distant sources together with that of the observed nearby SNRs listed in Table 1 is not able to explain the observed CR proton flux below $\sim 100 GeV$. It is unlikely that it can be explained by the inclusion of the SNRs G311.5-0.3 and Lupus Loop (whose ages are not known) in the analysis since the CR proton flux depends strongly on the source distance $r$ and both of them are at a distance of $r\sim 1.2 kpc$ away from us. From fig.2 it can be seen that the maximum proton flux from these two sources at $10 GeV$ energies will be approximately 3$-$4 orders of magnitude less than the observed flux. 
To explain the observed CR proton spectrum below $\sim 100 GeV$, the presence of at least one nearby source with a characteristic age of $t\sim 10^5yrs$ located at a distance of $r\sim 0.1kpc$ is essential under the assumption of a burst-like injection of particles at time $t=0$.  This \textit{missing} source may be any old SNR which is unable to be detected by the present radio/x-ray telescopes because of their limited sensitivities. In fact, assuming adiabatic phase the surface brightness estimated from an SNR of age $\sim 10^5$ yrs is fainter than the detection limit of the radio telescopes in studies of Galactic SNRs distribution (Kodaira 1974, Leahy $\&$ Xinji 1989). Such old SNRs can contribute a significant flux of CR protons in the solar vicinity although the electrons present inside it are not energetic enough to produce a strong radio/x-ray flux that can be detected by the present day telescopes. Incidentally, the parameters $(r,t)$ of the \textit{missing} SNR obtained in this work are found to be almost same with that obtained by Atoyan et al. 1995 in the study of local CR electrons where a nearby SNR is required to explain the observed high energy fluxes of electrons. 

Regarding the anisotropy that will be observed on Earth due to the \textit{missing} source, the degree of anisotropy can be calculated using the relation 
\begin{equation}
\xi_i = \frac{3}{2c}\frac{r_i}{t_i}
\end{equation} 
(Shen $\&$ Mao 1971) where $r_i$ and $t_i$ are the distance and age of the source respectively. For the source parameter obtained here i.e. $r_i\sim 0.1 kpc$ and $t_i\sim 10^5 yrs$, the degree of anisotropy of CR protons on Earth is found to be $\xi_i\sim 5\times 10^{-3}$ on the assumption of burst-like injection of protons at time $t=0$.

\end{document}